# Poster: Did I Just Browse A Website Written by LLMs?

Sichang Steven He
University of Southern California
Los Angeles, USA

Ramesh Govindan
University of Southern California
Los Angeles, USA

Harsha V. Madhyastha
University of Southern California
Los Angeles, USA

## ABSTRACT

Increasingly, web content is automatically generated by large language models (LLMs) with little human input. We call this "LLM-dominant" content. Since LLMs plagiarize and hallucinate, LLM-dominant content can be unreliable and unethical. Yet, websites rarely disclose such content, and human readers struggle to distinguish it. Thus, we must develop reliable detectors for LLM-dominant content. However, state-of-the-art LLM detectors are inaccurate on web content, because web content has low positive rates, complex markup, and diverse genres, instead of clean, prose-like benchmark data SoTA detectors are optimized for.

We propose a highly reliable, scalable pipeline that classifies entire websites. Instead of naively classifying text extracted from each page, we classify each site based on an LLM text detector's outputs of multiple prose-like pages to boost accuracies. We train and evaluate our detector by collecting 2 distinct ground truth datasets totaling 120 sites, and obtain 100% accuracies testing across them. In the wild, we detect a sizable portion of sites as LLM-dominant among 10k sites in search engine results and 10k in Common Crawl archives. We find LLM-dominant sites are growing in prevalence and rank highly in search results, raising questions about their impact on end users and the overall Web ecosystem.

## 1 BACKGROUND

Nowadays, AI services facilitate almost anyone to cheaply mass generate web content. AI blog generators can expand single sentences to entire blogs, while AI website generators take it further. For instance, through simple steps of clicking and answering questions, in minutes, and at no cost, Wix.com can generate entire, cohesive sites with boilerplate, blog, and service pages.

Users already complain about seeing LLM-dominant articles online; the EU AI Act even mandates the disclosure of AI use. However, websites rarely disclose generated content; and humans struggle to distinguish it [2].

To reliably detect LLM-dominant web content, we identify **3 key challenges**. *a) Text Detector Inaccuracies:* [2] has shown SoTA text detectors to be inaccurate at the low false positive rates (FPR) required for low-positive-rate real-world settings. We confirm such inaccuracies in §3. *b) Web Content Noise:* Detectors are designed for clean prose and perform poorly on the diverse genres of the Web, such as link listing or privacy notice. *c) Lack of Ground Truth:* There are many benchmark datasets for detecting text segments, but not for webpages, so it is difficult to evaluate web content detectors.

## 2 WEBSITE DETECTION PIPELINE

Our pipeline (Fig. 1) reliably detects LLM-dominant web content on the website level. It addresses the challenges above by acquiring clean, prose-like texts and aggregating a text detector's outputs for each site.

**Text Acquisition.** For each website, we 1) sample pages randomly, either directly from its sitemap if it exists, or from Wayback Machine Content Index's recent pages. Then, we 2) visit each page with Chromium and render its HTML, and 3) use Trafilatura [1] to extract the main textual content.

**Scoring and Filtering.** We detect LLM text with Binoculars [3]. While our pipeline is detector-agnostic, we choose Binoculars because it outperforms all 11 other detectors in our testing, and [2] shows that it excels under many topical domains, source LLMs, adversarial attacks, and low FPR. However, our initial experiments show Binoculars misclassifies many pages, including link listing, interactive interface, boilerplate like privacy notice, and even short text. We reason that LLM text detectors are designed for longer prose, not web content's diverse, noisy genres *(Web Content Noise)*. To reduce noise, we 4) apply strict filtering on each page's main content, discarding pages based on short texts, high ratios of text in lists, tables, or links, and intra-site text duplication. Consequently, most filtered texts are prose, enabling Binoculars to perform well.

**Aggregate Analysis.** To address *Text Detector Inaccuracies*, we assume pages from LLM-dominant website tend to be generated by similar means and form a distribution. For each page, Binoculars outputs a score indicating how likely a human wrote it; so, by aggregating Binoculars Scores from more pages into a sample distribution, we obtain a more robust signal for detection. Thus, for each site, we sample 15 to 20 pages through $1 \sim 4$), then 5) compute their Binoculars Scores. To capture each site's distribution, we compute the 9 deciles of its Binoculars Scores as the feature vector. Using these feature vectors, we 6) build a linear support vector machine (SVM) to classify whether a website is LLM-dominant.

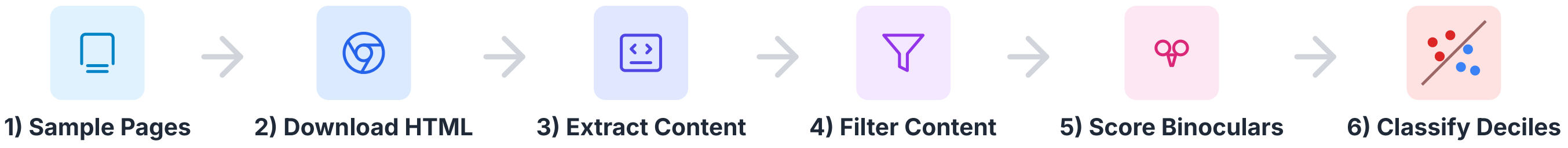

**Figure 1: Overview: LLM-Dominant Website Detection Pipeline.**

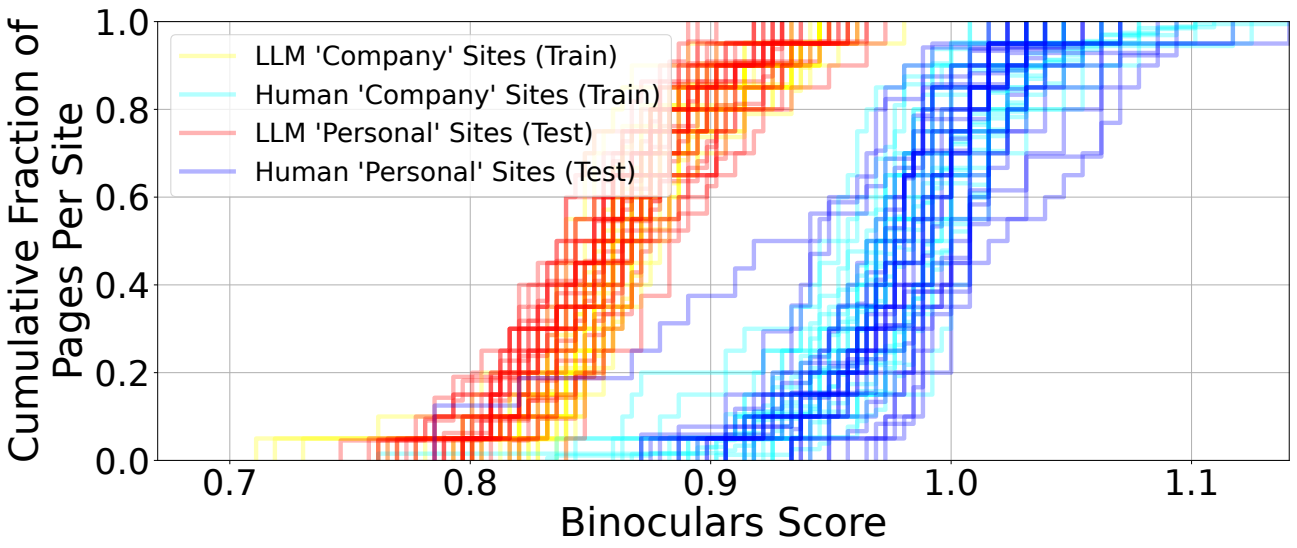

**Figure 2: CDFs of Binoculars Scores for Baselines Sites.**

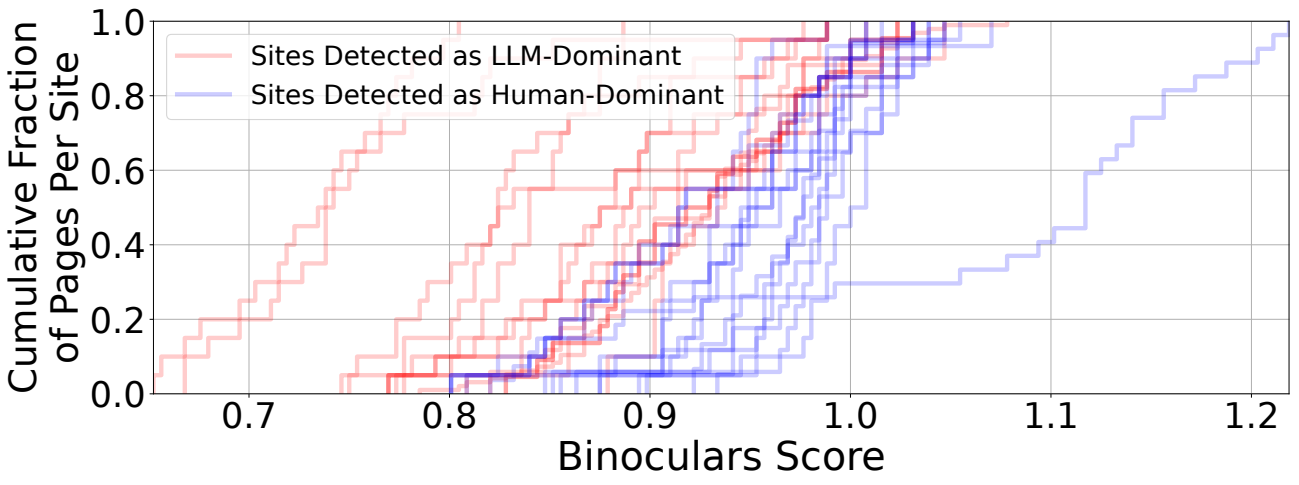

**Figure 3: CDFs of Binoculars Scores of Representative LLM, Human, and In-Between Search Result Sites.**

## 3 EVALUATION ON BASELINES

Facing the *Lack of Ground Truth* for training and testing our website detection pipeline, we constructed 2 baseline datasets totaling 120 sites, 2630 pages after filtering. For rigorous out-of-distribution (OOD) tests, the 2 datasets are from distinct sources. 'Company' dataset contains 30 human(-dominant) company sites of multiple sectors from the Russell 2000 stock index; then, with descriptions of each human site, we prompted Wix.com's AI website builder to generate 30 LLM(-dominant) sites, each with 20 blogs. Likewise, 'Personal' dataset includes 30 personal sites from IndieWeb Blogs, and 30 corresponding LLM sites generated by B12.io.

Our detector generalizes to 100% OOD accuracies across these distinct datasets, either when trained on 'Company' dataset and tested on 'Personal' (Fig. 2), or vice versa. Fig. 2 shows Binoculars Scores of many LLM and human pages overlap, in part due to remaining noise, such as the privacy policy pages Wix generated. Since Binoculars classifies with a threshold, its per-page accuracy is at best 93%. Despite that, the SVM still completely separates the LLM sites, exemplifying the robustness of aggregate analysis against noise.

## 4 FINDINGS IN THE WILD

We trained our detector on both baselines and ran it on 2 sets of websites in the wild to study how frequently users may face LLM-dominant sites.

**Search Engine Results.** To estimate how much LLM(-dominant) web content users may see, we searched 2000 WikiHow article titles, got 17,036 sites in the top 20 results from Bing[1], and ran our pipeline over them. Many sites did not yield 15 pages, mainly due to permission and connectivity issues. Of the 10,232 sites with at least 15 pages after filtering (264,918 pages total), our pipeline detected a sizable 1019 as LLM (9.96%). These sites often seem to be generic blogs with many ads, written by made-up authors. Fig. 3 illustrates the Binoculars Score distributions of representative sites. Unlike our baselines (Fig. 2), we see a blurry boundary between LLM and human sites, which suggests there are many in-between sites partially composed of LLM-dominant pages. Moreover, we find no statistically significant difference that indicates LLM sites rank lower in search results (equal medians; means within $10.66 \pm 0.035$), which suggests Bing is not effectively penalizing LLM-dominant content, potentially degrading user experience.

**Common Crawl.** To understand the historical trend, we analyzed 10,479 random sites across Common Crawl archives from 2020 to 2025 (284,523 pages). Overall, only 451 sites (4.30%) are detected as LLM(-dominant), much lower than the 9.84% in search results, suggesting web search surfaces much more LLM sites than randomly sampling the Web. However, of the 4938 sites crawled entirely after ChatGPT launched, a sizable 358 (7.25%) are LLM; of the 764 sites started being crawled between 2024 and 2025, an even larger 77 (10.08%) are LLM. This indicates the portion of LLM-dominant sites is growing. Finally, of the 1315 sites crawled entirely before ChatGPT launched, 16 (1.22%) are misclassified as LLM, mainly due to pages evading our filters and can be eliminated; this shows our detector has low but non-zero FPR.

## 5 FUTURE WORK

We will investigate existing challenges, like false positives and in-between sites in §4, and better evaluation methods; we also see lots of green fields around LLM-dominant web content. While some of these sites seem to farm ads for revenue, in general, we want to understand who generates such content, and why and how they do so. In extension, we want to detect AI images and other AI content. Finally, we would like to quantify the impact of such content on the Web ecosystem, such as on the quality of search engines results.

## A ETHICS

Our work does not raise any ethical concerns, because all data we obtained are from the public web and sites we generated are generally unreachable.

[1]Bing's API returns results consistent with its web interface.

## REFERENCES

[1] Adrien Barbaresi. 2021. Trafilatura: A Web Scraping Library and Command-Line Tool for Text Discovery and Extraction. In *ACL*.
[2] Liam Dugan et al. 2024. RAID: A Shared Benchmark for Robust Evaluation of Machine-Generated Text Detectors. In *ACL*.
[3] Abhimanyu Hans et al. 2024. Spotting llms with binoculars: Zero-shot detection of machine-generated text. In *ICML*.